%% file: 28417_ap.tex
\documentclass[letter]{aa} 

\usepackage{graphicx}
\usepackage{txfonts}
\usepackage{hyperref}
\usepackage{color}

\def\Mjup{\hbox{$\thinspace M_{\mathrm{J}}$}}
\def\Msun{\hbox{$\thinspace M_{\odot}$}}
\def\Rsun{\hbox{$\thinspace R_{\odot}$}}
\def\Lsun{\hbox{$\thinspace L_{\odot}$}}
\def\Teff{\hbox{$\thinspace T_{\mathrm{eff}}$}}
\def\Teq{\hbox{$\thinspace T_{\mathrm{eq}}$}}
\def\kms{\hbox{$\thinspace {\mathrm{km~s^{-1}}}$}}
\def\ms{\hbox{$\thinspace {\mathrm{m~s^{-1}}}$}}
\def\amag{\hbox{$\thinspace \mathrm{mag}$}}
\def\ALi{\hbox{$\thinspace A (\mathrm{Li})$}}
\def\K{\hbox{$\thinspace \mathrm{K}$}}

\def\au{\hbox{$\thinspace \mathrm{au}$}}
\def\sjit{\hbox{$\sigma_{\mathrm{jitter}}$}}
\def\srv{\hbox{$\sigma_{\mathrm{RV}}$}}
\def\sbs{\hbox{$\sigma_{\mathrm{BS}}$}}
\def\Vosc{\hbox{$v_{\mathrm{osc}}$}}
\def\Vrot{\hbox{$v_{\mathrm{rot}}$}}
\def\Posc{\hbox{$P_{\mathrm{osc}}$}}
\def\Prot{\hbox{$P_{\mathrm{rot}}$}}
\def\Umib{\hbox{$8\thinspace\mathrm{U}\thinspace\mathrm{Mi}\thinspace\mathrm{b}$}}

\begin{document}

  \title{Tracking Advanced Planetary Systems (TAPAS) with HARPS-N.  
  \thanks{Based on observations obtained with the Hobby-Eberly Telescope, which is a joint project of the University of Texas at Austin, the Pennsylvania State University, 
Stanford University, Ludwig-Maximilians-Universit\"at M\"unchen, and Georg-August-Universit\"at G\"ottingen.}
\thanks{Based on observations made with the Italian Telescopio Nazionale Galileo (TNG) operated on the island of La Palma by the Fundaci\'on Galileo Galilei of the INAF (Istituto Nazionale di Astrofisica) at the Spanish Observatorio del Roque de los Muchachos of the Instituto de Astrof\'{\i}sica de Canarias.}
}

   \subtitle{IV. TYC 3667-1280-1 -  the most massive red giant 
   star hosting a warm 
   Jupiter. }

   \titlerunning{TAPAS IV. TYC 3667-1280-1 b -  the most massive red giant star hosting a warm Jupiter. } 
   \authorrunning{A. Niedzielski et al.}

 \author{A. Niedzielski
          \inst{1}  
                              \and
          E. Villaver
         \inst{2}
                \and
         G. Nowak
          \inst{3,4,1}
                                                \and                                
M. Adam\'ow
          \inst{5,1}
\and
         G. Maciejewski
          \inst{1}
          \and
K. Kowalik
          \inst{6}         
          \and                                                             
          A. Wolszczan
          \inst{7,8}
                                      \and
         B. Deka-Szymankiewicz
          \inst{1}
                                       \and
                                     M. Adamczyk
          \inst{1}
                   }
                   
   \institute{Toru\'n Centre for Astronomy, Faculty of Physics, Astronomy and Applied Informatics, Nicolaus Copernicus University in Toru\'n, Grudziadzka 5, 87-100 Toru\'n, Poland.
              \email{Andrzej.Niedzielski@umk.pl}
                                    \and
        Departamento de F\'{\i}sica Te\'orica, Universidad Aut\'onoma de Madrid, Cantoblanco 28049 Madrid, Spain.
         \email{Eva.Villaver@uam.es}
             \and
              Instituto de Astrof\'isica de Canarias, E-38205 La Laguna, Tenerife, Spain.
              \and
              Departamento de Astrof\'isica, Universidad de La Laguna, E-38206 La Laguna, Tenerife, Spain.
              \and
              McDonald Observatory and Department of Astronomy, University of Texas at Austin, 2515 Speedway, Stop C1402, Austin, Texas, 78712-1206, USA.
\and
National Center for Supercomputing Applications, University of Illinois, Urbana-Champaign, 1205 W Clark St, MC-257, Urbana, IL 61801, USA                  
         \and
             Department of Astronomy and Astrophysics, Pennsylvania State University, 525 Davey Laboratory, University Park, PA 16802, USA.
          \email{alex@astro.psu.edu}
         \and
             Center for Exoplanets and Habitable Worlds, Pennsylvania State University, 525 Davey Laboratory, University Park, PA 16802, USA.
             }

   \date{Received;accepted}

 
  \abstract
   { We present the latest result of the TAPAS project that is
devoted to intense monitoring of planetary candidates that are identified within the PennState-Toru\'n planet search. }
   {We aim to detect planetary systems around evolved stars to be able to build sound 
   statistics on the frequency and intrinsic nature of these systems, and to deliver in-depth 
   studies of selected planetary systems with evidence of star-planet interaction processes.}
   {The paper is based on precise radial velocity measurements: 13 epochs 
   collected over 1920 days with the Hobby-Eberly Telescope and its High-Resolution Spectrograph, and 22 epochs of ultra-precise HARPS-N data collected over 961 days. }
   {We present a warm-Jupiter ($\Teq=1350\K$, $m_2\sin i=5.4\pm0.4\Mjup$)  companion with
   an orbital period of 26.468 days in a circular ($e=0.036$) orbit around a giant
   evolved ($\log g=3.11\pm0.09$, $R=6.26\pm0.86\Rsun$) star with
   $M_{\star}=1.87\pm0.17  \Msun$. This is the most massive and oldest star found
   to be hosting a close-in giant planet. Its proximity to its host ($a=0.21\au$)
   means that the planet has a $13.9\pm2.0\%$ probability of transits; this calls for
   photometric follow-up study. }
   {This massive warm Jupiter with a near circular orbit around an evolved massive star can help set constraints on general migration mechanisms for warm Jupiters and, given its high equilibrium temperature, can help test energy deposition models in hot Jupiters.}


   \maketitle
%

\section{Introduction}

The population of hot Jupiters (HJ), Jupiter-mass planets on short-period orbits ($P<10$ days, or  within 0.1\au),
which were unveiled with the finding of 51 Peg b \citep{Mayor1995}, was probably one of the least expected discoveries
in search for exoplanets. The origin of this rare class of objects, only known to be present around approximately 0.5--1\% of FGK stars \citep{Wright2012, Howard2012},
generally involves early (e.g., Type II), or late migration scenarios (following planet-planet scattering or secular perturbations from more distant objects) since formation beyond the snow line is required. Recently, however, the in-situ formation of these systems has also been proposed \citep{Boley2016}.
\par Warm Jupiters (WJ, \citealt{Fogg2009}), Jupiter-mass planets on 10--100 day orbits, are even more intriguing. WJ have observed eccentricities too low for significant tidal evolution, and while the missing HJ from the evolved star-planet population can be naturally explained by engulfment by the star, the lack so far
of detected WJ requires further explanation (see, e.g., \citealt{Villaver2014}). Recently, \cite{Frewen2016} invoked Kozai-Lidov oscillations to predict an entire removal of WJ planets
by the time the star reaches $R>5\Rsun$ , while an identical constant eccentricity population survives beyond $40\Rsun$ .
The authors argued that the WJ migrate through Kozai-Lidov oscillations. \citet{Dawson2013} presented evidence that both smooth disk migration and dynamical interactions may play a role in WJ formation, with the efficiency depending on the host's metallicity. \citet{Dong2014} showed that high-eccentricity WJ are more likely to have an external Jovian perturber and migrate through dynamical interactions.  
This finding was recently supported by \citet{Huang2016}, who found that different from HJ, half of the WJ have companions.
The mutual inclination of orbits in six such WJ-external perturber pairs reported by \citet{Dawson2014} indeed points to Kozai-Lidov oscillations as a mechanism responsible for the inward migration of WJ.

We here present  the discovery of a warm-Jupiter planet (P = 26.468 d) orbiting an evolved star with $R= 6.3 \Rsun$
on a circular orbit. Clearly, this unique system can help set constraints on the invoked mechanism for WJ formation through the 
Kozai-Lidov oscillations. Furthermore, its proximity to a very luminous star means that it has the equivalence temperature of a hot Jupiter.  If hot-Jupiter inflation works by depositing irradiation into the planet's deep interiors, then planetary radii should increase in response to the increased irradiation of this evolved star, constraining structure theories on the physics of the hot-Jupiter inflation mechanism \citep{Lopez2016}. Moreover, there is a high probability of transit detection of this unique object,
which may then serve as a perfect laboratory for planet formation and migration theories.

The paper is organized as follows: in
Sect. \ref{observations} we present the available  observations  and  discuss the influence of the stellar activity
on the radial velocity (RV) variation measurements.  Section \ref{results-g} shows the results of
the Keplerian data modeling, in Sect. \ref{transits} we discuss the transit possibilities, and in Sect.
 \ref{conclusions} we discuss the results  and present the conclusions.


\section{Observations, radial velocities, and activity\label{observations}}

TYC 3667-1280-1 (2MASS J00513296+5825342) belongs to a sample of  about 300 planetary or brown dwarf (BD) candidates 
identified in the complete sample of over 1000 stars for RV
variations with the 9.2 m Hobby-Eberly Telescope  (HET, \citealt{Ramsey1998})
and its High-Resolution
Spectrograph   (HRS,
\citealt{Tull1998}) since 2004 within the {PennState - Toru\'n Centre
for Astronomy Planet Search} (PTPS, \citealt{Niedzielski2007, Niedzielski2016a})   
selected for a more intense precise RV follow-up within 
 {Tracking
Advanced Planetary Systems (TAPAS) with HARPS-N}   \citep{Niedzielski2015a, Adamow2015,Niedzielski2016b}.
A summary of the available data for TYC 3667-1280-1 is given in Table \ref{Parameters1}.

\begin{table}
\centering
\caption{Summary of the available data on TYC 3667-1280-1.}
\input{table1s}

\tablefoot{References: (1)~\cite{Hog2000}, (2)~\cite{Zielinski2012}, (3)~\cite{Nowak2012}, (4)~\cite{Adamow2014},
        (5)~\cite{Adamczyk2015}, (6)~calculated from M$_{V}$, (7)~This work.}
\label{Parameters1}
\end{table}

The spectroscopic observations presented here were made  with
the HET 
 and its  HRS in the queue-scheduled mode \citep{Shetrone2007}
and with the 3.58 m Telescopio Nazionale Galileo (TNG) and its High-Accuracy Radial velocity Planet Searcher in the Northern hemisphere (HARPS-N,
\citealt{Cosentino2012}). 

For  HET HRS spectra we  used a combined gas-cell  \citep{MarcyButler1992, Butler1996}, and  cross-correlation
\citep{Queloz1995, Pepe2002} method for precise RV and spectral line bisector (BS) measurements, respectively.  
The implementation of this technique to our data is described in \cite{Nowak2012} and \cite{ Nowak2013}. 

HARPS-N radial velocity measurements and their uncertainties as well as BS were obtained
with the standard user pipeline, which is based on the weighted
CCF method \citep{1955AcOpt...2....9F, 1967ApJ...148..465G, 1979VA.....23..279B, Queloz1995, Baranne1996, 
  Pepe2002}, using the simultaneous Th-Ar calibration 
mode of the spectrograph  and the K5 
cross-correlation mask. 
The RV and BS data for TYC 3667-1280-1 are presented in Tables \ref{HETdata} and \ref{HARPSdata}.

There is no correlation between RV and BS  in either HET (Pearson's $r=0.07$) or TNG ($r=-0.19$) data, and we can assume that the RV signal origin is Dopplerian.
Our H$\alpha$ index \citep{2013AJ....146..147M} variations are neither
correlated with the observed RVs  ($r=-0.17$) nor with the control line FeI
$6593.878\thinspace\AA$ ($r=0.37$) and therefore show no trace of stellar
activity.
The Ca~II H and K lines  present in the TNG HARPS-N spectra allow us to 
calculate the instrumental S$_{\mathrm{HK}}^{\mathrm{inst}}$ index  \citep{Niedzielski2015a}, which is
$0.17 \pm 0.08$ and indicates that  TYC 3667-1280 is  not an active star. There are no emissions in the core of H and K lines. 

Extensive photometry for TYC 3667-1280-1 is available from NSVS \citep{2004AJ....127.2436W} and WASP \citep{Pollacco2006}.
The 157 epochs of NSVS photometry collected between  JD 2451336.35 -- 2451566.14 show constant light 
$m_{\mathrm{NSVS}}=9.66\pm0.03\amag$ and no trace of variability. 1833 epochs of WASP photometry, obtained 
from the original 2064 through three-sigma iterative filtering were collected between JD 2454327.73 -- 2454452.35
and are partly contemporaneous with our HET HRS data. They show $m_{\mathrm{WASP}}=9.927 \pm0.009\amag$,
varying within $0.022\amag$ and a trace of weak periodic signal at 15.67 days. This signal is well separated from 
that present in RV and also too short to be related to stellar rotation and a possible spot on the stellar surface.

\section{Keplerian analysis \label{results-g}}

\begin{figure}
   \centering
   \includegraphics[width=0.5\textwidth]{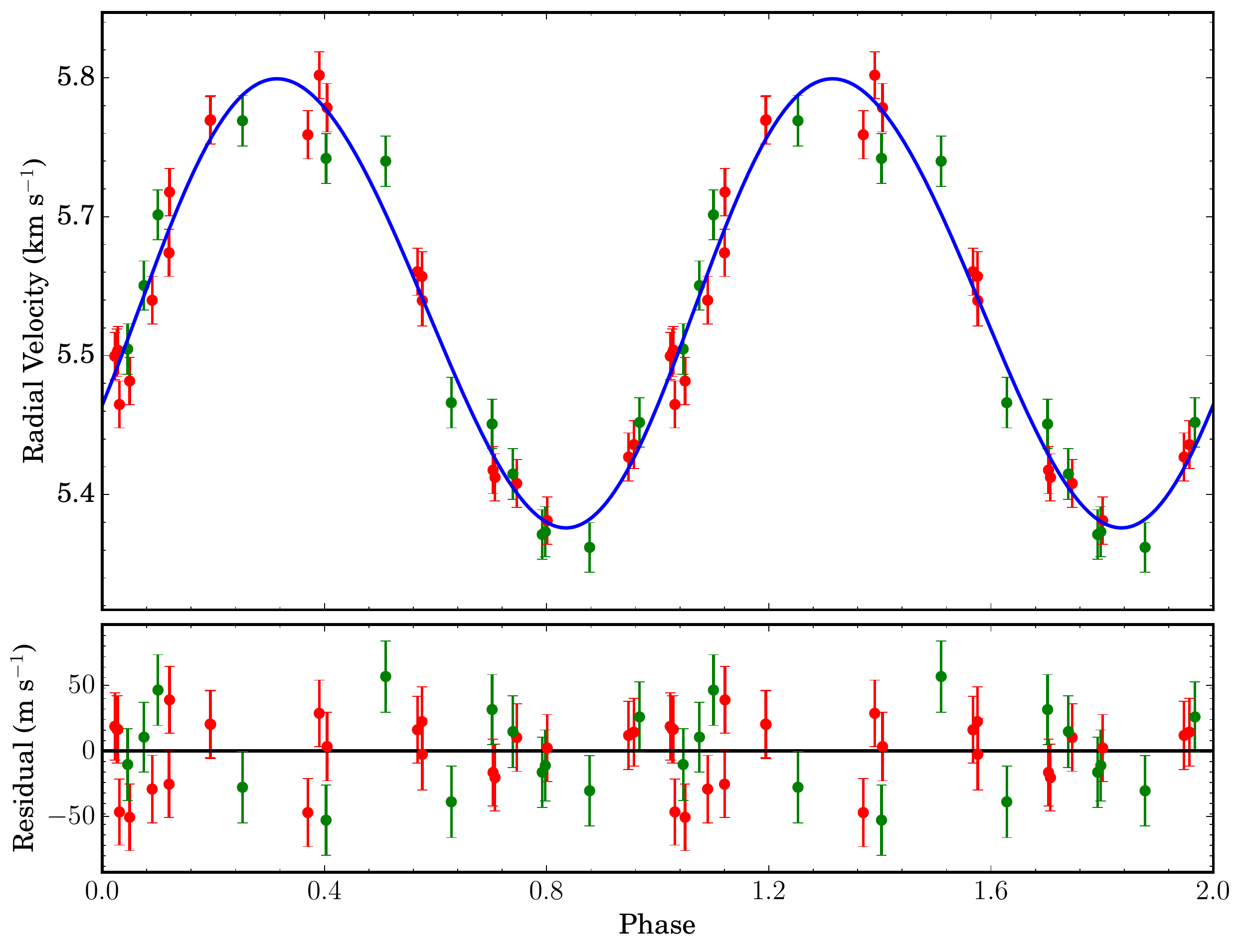}
   \caption{Keplerian best fit to combined HET HRS (green points) and TNG Harps-N (red points) data for TYC 3667-1280-1. The jitter is added to uncertainties.}
   \label{Fit_1}
\end{figure} 

Keplerian orbital parameters  were derived using a hybrid approach (e.g., \citealt{2003ApJ...594.1019G, 2006A&A...449.1219G, 2007ApJ...657..546G}), 
in which  the PIKAIA-based, global genetic algorithm (GA; \citealt{Charbonneau1995})  was combined with 
MPFit algorithm \citep{Markwardt2009} to find the best-fit Keplerian orbit  delivered by RVLIN \citep{WrightHoward2009}
modified  to allow the stellar jitter to be fitted as a free parameter \citep{2007ASPC..371..189F, 2011ApJS..197...26J}.
The RV bootstrapping  method  \citep{1993ApJ...413..349M, 1997A&A...320..831K, Marcy2005,
  Wright2007}  was employed to assess the uncertainties of the best-fit orbital parameters (see \cite{Niedzielski2015a} for more details).
The results of the Keplerian analysis are presented in Table \ref{KeplerianFits} and Fig. \ref{Fit_1}.

\begin{table}
\centering
\caption{Keplerian orbital parameters of TYC 3667-1280-1 b.}
\input{table_fit2}
\tablefoot{{ $v_0$ denotes absolute velocity of the barycenter of the system,
offset is a shift in radial velocity measurements between different telescopes,
\sjit~is stellar intrinsic jitter as defined in \cite{2011ApJS..197...26J},
RMS~is the root mean square of the residuals.}}
\label{KeplerianFits}
\end{table}

\section{Possibility of transits \label{transits}}

The architecture of the system, with the relatively short orbital period of 26.468 days and the semi-major axis
of only $7.2 \pm 1.0$ stellar radii, results in a probability
of $13.9\pm2.0$\% that the planet transits the host star.

{ The star is expanding because of its evolution toward the giant branch, which means that the planet is expected to be bloated, with the equilibrium temperature $T_{\rm{eq}}=1350 \pm 100$ K. Assuming that the planetary radius is in a range of between 1 and 2 Jupiter radii, the transits are expected to be 0.3--1.1 mmag deep. The predicted transit duration is between 5.2 h for grazing transits and 29 h for central transits. These transit parameters together with possible photometric variability of the host star make transit observations challenging from the ground with telescopes smaller than 2m. The system could be followed with space-borne instruments such as CHEOPS \citep{2013EPJWC..4703005B}.
The ephemeris of  the possible transits is 
\begin{equation}
  T_{\rm{t}} (\rm{HJD}) = 2457235.4 \pm 2.6 + E \times P\, , \;
\end{equation} 
where $T_{\mathrm{t}}$ is the mid-transit time for an epoch $E$ (in HJD).}
 
Detecting a transit event would allow determining the planetary radius, orbital inclination, planetary mass, and mean density .
The planet would be a very interesting object for studies of planet-star interactions in late stages of stellar evolution.
A non-detection would set constraints on the orbital inclination or planetary radius.  

\section{Discussion and conclusions\label{conclusions}}

Given the orbital period of the companion (26.468 days) and the stellar
mass  ($M=1.87\pm0.17 \Msun$), stellar radius ($R = 6.26\pm0.86 \Rsun$), and the evolutionary stage of the host ($\log g=3.11\pm0.09$), the planetary system TYC
3667-1280-1 is certainly  unique.  Its host is an evolved A star entering the
red giant branch.  
The planet resides at only
$7.2\pm1.0\thinspace R_{\star}$ and is on the brink of engulfment, which will
occur when its host reaches the radius of $\sim 20\Rsun$ \cite{Villaver2014}
in $\sim 1.7\times10^8\thinspace\mathrm{yr}$. This fate is only marginally
delayed using a different mass-loss prescription (see \citealt{Villaver2014}). 

\begin{figure}
   \centering
   \includegraphics[width=0.46\textwidth]{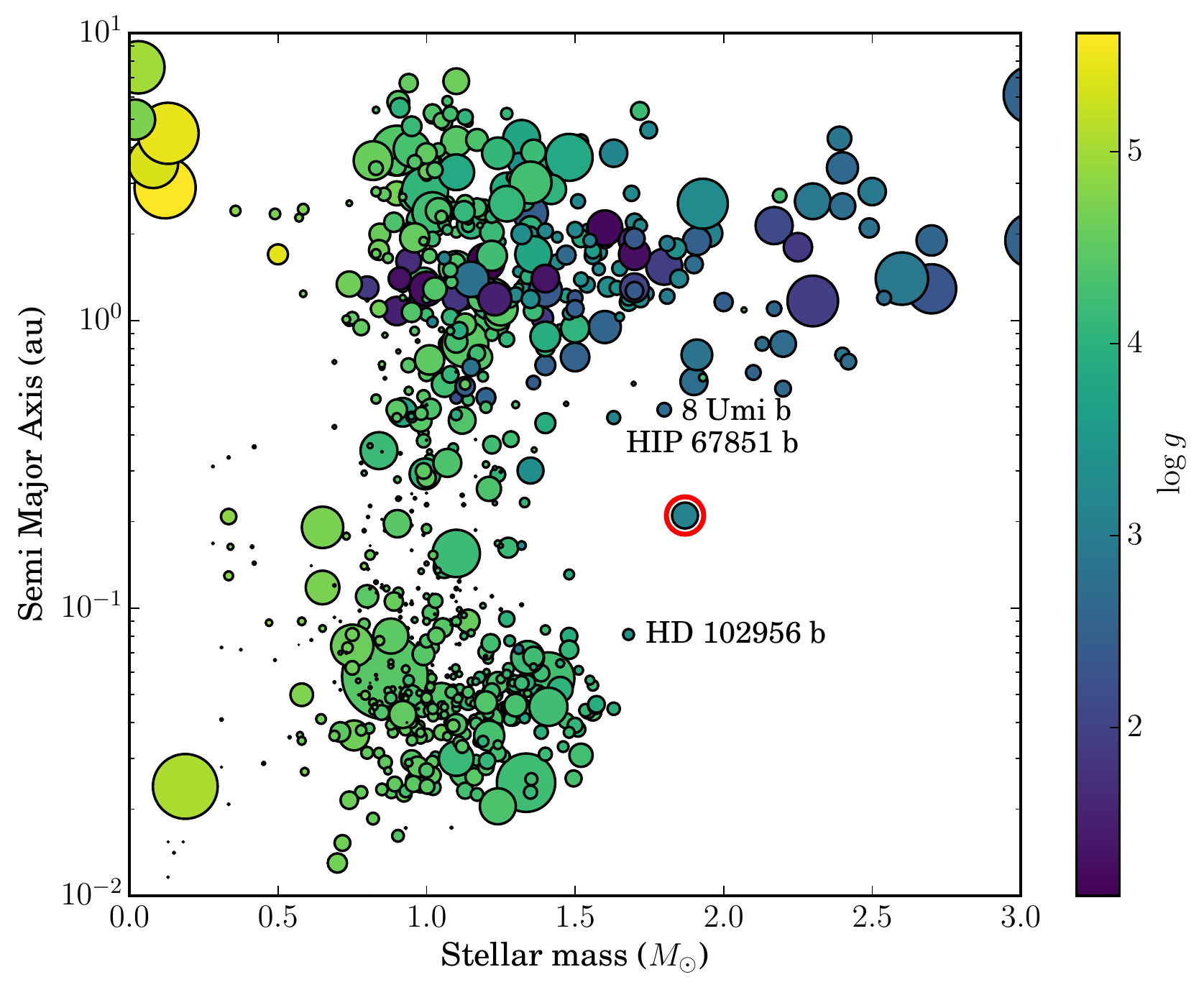}
   \caption{Orbital period vs. stellar radius plot with the position of TYC
   3667-1280-1 b indicated with a red circle. The size of the symbols is proportional
   to the planetary mass, color indicates $\log g$ of the host star. Data from
   \href{http://www.exoplanet.eu/catalog/csv/}{Exoplanets.eu}.}
   \label{mass-a}
\end{figure}

It is the most compact planetary system ($a=0.21\au$) hosted by an evolved
($\log g\leq3.5$), intermediate-mass ($M_{\star}\geq 1.5 \Msun$) star.  The
only planetary systems discovered using RV method that bear a resemblance to
TYC 3667-1280-1 are HIP~67851~b \citep{Jones2015}, HD~102956~b
\citep{Johnson2010}, and \Umib~\citep{2015A&A...584A..79L}.  However, while the host star of HIP
67851 b has a similar mass of $1.63\pm0.22\Msun$ and is at the
similar stage of stellar evolution ($\log g=3.2\pm0.2$), the planet itself
resides on a much wider orbit with $a=0.459\pm0.021\au$ and is much lighter
($1.38\pm0.15\Mjup$).  HD 102956 b is even less massive ($0.96\pm0.05\Mjup$)
and is hosted by a less evolved ($\log g=3.5\pm0.06$), $1.68\pm0.11 M_{\odot}$
star, similarly to \Umib, a $1.5\pm0.2\Mjup$ planet orbiting a
$1.8\pm0.1\Msun$, $\log g=2.57\pm0.03$ star in $0.49\pm0.03\au$ orbit (Fig.
\ref{mass-a}).

\begin{figure}
   \centering
   \includegraphics[width=0.46\textwidth]{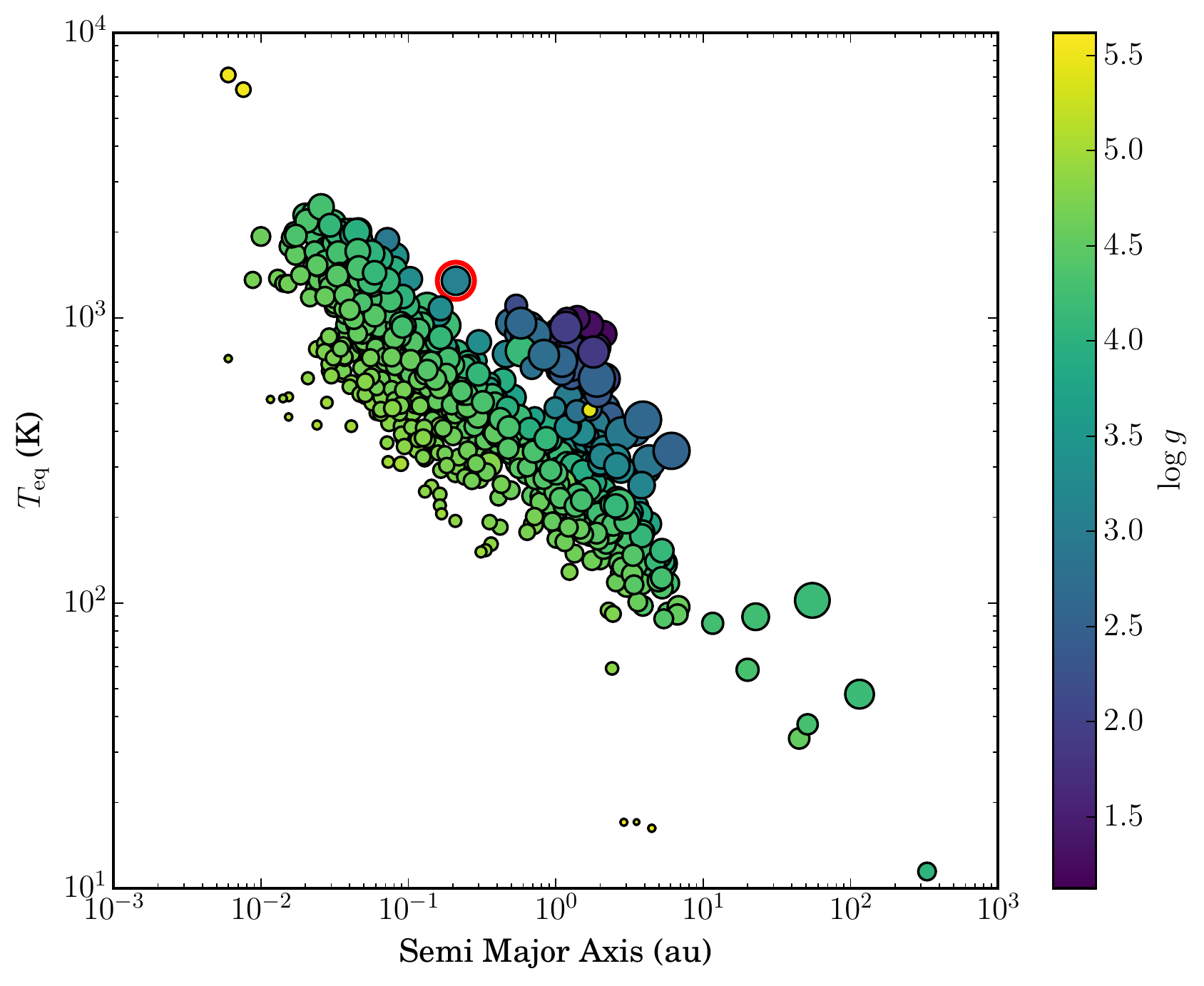}
   \caption{Planet equilibrium temperature vs. semi-major axis plot with
   the position of TYC 3667-1280-1 b indicated with a red circle. The size of the symbols
   is proportional to the stellar mass, color indicates $\log g$ of the host
   star.  Data from \href{http://www.exoplanet.eu/catalog/csv/}{Exoplanets.eu}.}
   \label{Teq}
\end{figure}

TYC 3667-1280-1 b is the most massive planet of these, which makes it a strong
case for the study of tidal dissipation, similar to WASP-18b
\citep{Hellier2009}. The proximity of its evolved luminous host
TYC~3667-1280-1~b causes its equilibrium temperature to be very high and places the planet
among the hottest Jupiters known (see Fig.~\ref{Teq}). This makes it a perfect
laboratory for studying inflation processes \citep{Lopez2016}.

The relatively high probability of transits adds to is attractiveness. The
estimated probability of $13.9\pm2.0$\% is high enough to make this worthwhile, even though
the expected transit depth is relatively low (0.3--1.1 mmag) and the transit
duration may reach up to 29 hours. TYC 3667-1280-1~b certainly deserves more
attention and a photometric follow-up in search for transits.

Furthermore, the planetary system of TYC 3667-1280-1 represents an
excellent test case on which to study the formation of WJ through Kozai-Lidov oscillations
because its mere existence, along with its almost circular orbit, defies the
theories of migration through this mechanism. Planets like TYC 3667-1280-1~b are
extremely rare. One object in the complete sample of $\sim1000$ PTPS stars
suggests a frequency of only $\sim0.1\%$.  However, given one object detected so
far within the total PTPS  sample of 
103 stars with $M_{\star}\geq 1.5\Msun$ , we can estimate that objects like this
appear around $\sim 1\%$ of evolved intermediate-mass  stars in our sample.
They are apparently as common as HJ around the main-sequence stars.  Our
discovery suggests that WJ around intermediate-mass stars survive the
evolution of their hosts at the beginning of the red giant branch, where TYC
3667-1280-1 is currently located.

\begin{acknowledgements}
We thank the HET  and IAC resident astronomers and telescope operators  for
their support.

MA acknowledges the Mobility+III fellowship from the Polish Ministry of Science
and Higher Education. 

MA, AN, BD, and MiA  were supported by the Polish National Science Centre grant
no. UMO-2012/07/B/ST9/04415.  

EV acknowledges support from the Spanish Ministerio de Econom\'ia y
Competitividad under grant AYA2014-55840P.

KK was funded in part by the Gordon and Betty Moore Foundation's
Data-Driven Discovery Initiative through Grant GBMF4561.

This research was supported in part by PL-Grid Infrastructure.

The HET is a joint project of the University of Texas at Austin, the
Pennsylvania State University, Stanford University, Ludwig-
Maximilians-Universit\"at M\"unchen, and Georg-August-Universit\"at
G\"ottingen. The HET is named in honor of its principal benefactors, William P.
Hobby and Robert E. Eberly.  The Center for Exoplanets and Habitable Worlds is
supported by the Pennsylvania State University, the Eberly College of Science,
and the Pennsylvania Space Grant Consortium.

This work made use of NumPy~\citep{numpy}, Matplotlib~\citep{mpl},
Pandas~\citep{pandas} and \texttt{yt}~\citep{yt}.

\end{acknowledgements}

\bibliographystyle{aa} 
\bibliography{an2-EV.bib} 
\newpage

%
\begin{table}
\centering
\caption{HET and HRS RV and BS measurements  (\!\ms) of TYC 3667-1280-1}
\begin{tabular}{lrrrr}
\hline
MJD& RV  & \srv & BS  & \sbs \\
\hline\hline 
54394.155075  &    258.5  &     10.3    &  -48.4    &   14.4     \\
54398.137124   &   217.8  &     9.3   &  -36.9     &  24.0    \\
 54748.181551   &   -45.9   &    10.5   &   -20.8    &   31.5    \\
54779.129988   &  -185.2   &    10.0   &   -40.3    &   26.9    \\
55184.154144   &   156.9   &     9.8    &  -44.6    &   24.1    \\
55469.395590  &   -202.0   &     8.9    &  -83.8   &    19.7   \\
55500.319306    &   12.1    &   10.6    &   17.5    &   34.0    \\
55895.253142   &   -67.2   &     8.8    &  -28.3    &   21.9    \\
56127.428623  &   -122.8    &   10.6   &   -79.8   &    22.8    \\
56232.306667   &   -68.9   &     8.7     &  -0.4     &  19.4    \\
56280.170359   &   214.9  &     10.2    &  -90.6   &    27.1    \\
56295.121470  &     80.6    &    8.7  &   -10.7     &  27.9    \\
56314.096829   &  -188.1   &     9.2   &   -66.5   &    24.8    \\

\hline
\end{tabular}
\label{HETdata}
\end{table}
%
%
%

%
%
\begin{table}
\centering
\caption{TNG and HARPS-N RV and BS measurements (\!\ms) of TYC 3667-1280-1}
\begin{tabular}{lrrr}
\hline
MJD& RV & \srv & BS  \\
\hline\hline
56277.0056838   &      5852.76  &     0.52   &     -208.5    \\
56293.9512124  &        5497.29   &      0.89  &        -83.2    \\
56320.904931    &     5522.61   &    1.44   &     -63.3     \\
56560.1855872   &      5609.82  &      1.86   &     -150.9     \\
56647.0042254  &       5788.52   &        2.42   &     -70.0     \\
56684.8799384   &      5372.04   &        2.27  &       30.2     \\
56895.1655851   &      5412.09   &     2.56  &      -98.3     \\
56926.9657616  &       5440.58    &       2.93    &     19.5     \\
56927.2172308  &       5453.96   &      2.99   &      16.2     \\
56969.8405909   &      5640.65   &        1.84    &    -137.5     \\
56970.050176     &      5635.57    &       3.56    &    -187.8     \\
56970.070991    &     5609.47     &      4.10    &     -168.5      \\
56991.9991155   &      5817.72    &         2.75    &    -154.6     \\
57034.819799     &    5549.53     &       2.58   &      53.8     \\
57034.9211894   &      5553.18    &       2.77    &     57.9      \\
57034.9663153   &      5555.87    &        2.78   &      65.6      \\
57065.8310694   &      5803.94    &      1.99    &     16.8     \\
57065.8540671   &      5804.62    &         3.24    &     14.9     \\
57196.209713    &     5661.05     &        1.55   &      -95.9     \\
57196.2327686   &      5726.66    &        2.09   &      141.2     \\
57238.1151227   &      5426.42    &      1.63   &     -103.7     \\
57238.2091972   &      5418.56    &        1.79   &     -60.3     \\

\hline
\end{tabular}
\label{HARPSdata}
\end{table}

\end{document}

%% file: table1s.tex
{\small
\renewcommand{\arraystretch}{1.2}
\setlength{\tabcolsep}{5pt}
\begin{tabular}{ll|ll}
\hline
Parameter & Value & Parameter & Value \\
\hline
\hline
V  (mag)& $9.86 \pm 0.02^{(1)}$   & $M/\!\Msun$ & $1.87 \pm 0.17^{(5)}$  \\
B-V (mag) & $1.00 \pm 0.06^{(1)}$ & $\log L/\!\Lsun$ & $1.38 \pm 0.10^{(5)}$ \\
(B-V)$_0$ (mag) & $0.84^{(2)}$    & $R/\!\Rsun$ & $6.26 \pm 0.86^{(5)}$ \\
M$_V$ (mag) & $0.99^{(2)}$        & $\log \mathrm{age}$ (yr)& $9.13 \pm 0.10^{(5)}$ \\
$d$ (pc) & $481\pm 37^{(6)}$      & \Vosc (\!\ms) & $3.00^{+1.41}_{-0.95}\,^{(7)}$ \\
$\Teff$ (K) & $5130 \pm 24^{(2)}$ & \Posc (d) & $0.075^{+0.032}_{-0.024}\,^{(7)}$ \\
$\log g$ & $3.11 \pm 0.09^{(2)}$  & \Prot (d) & $99 \pm 19^{(7)}$ \\ 
$[$Fe/H$]$ & $-0.08 \pm 0.05^{(2)}$ & $\Vrot\sin i_{\star}$ (\!\kms) & $3.2 \pm 0.4^{(4)}$ \\
RV (\!\kms) & $9.54 \pm 0.06^{(3)}$ & $\ALi_{\mathrm{NLTE}} $& $0.80 \pm 0.21^{(4)}$ \\
\hline
\hline
\end{tabular}
\renewcommand{\arraystretch}{1}
\setlength{\tabcolsep}{6pt}
}

%% file: table_fit2.tex
{\small
\renewcommand{\arraystretch}{1.2}
\begin{tabular}{ll|ll}
\hline
Parameter & Value  & Parameter & Value \\
\hline
\hline
$P$ (days)             & $26.468^{+0.005}_{-0.005}$ & $v_0$ (\!\ms)  & $5610.4^{+2}_{-2.2}$ \\
$T_0$ (MJD)            & $56319.6^{+3.7}_{-3.3}$    & offset (\!\ms) & $5545^{+11}_{-11}$ \\
$K$ (\!\ms)            & $242.4^{+1.0}_{-1.2}$      & \sjit (\!\ms)  & $25.18$ \\
$e$                    & $0.036^{+0.043}_{-0.014}$  & $\sqrt{\chi_\nu^2}$    & $1.17$    \\
$\omega$ (deg)         & $240^{+50}_{-50}$ 	    & RMS (\!\ms)           & $28.3$    \\
$m_2\sin i$ (\!\Mjup)  & $5.4 \pm 0.4$ 	            & $N_{\textrm{obs}}$     & $35$      \\
$a$ (\!\au)            & $0.21 \pm 0.01$ 	    & & \\
\hline
\end{tabular}
\renewcommand{\arraystretch}{1}
}